# YOLOv5s-BC: An improved YOLOv5s-based method for real-time apple detection

Jingfan Liu[1], Zhaobing Liu[1,*]

[1] Hubei Digital Manufacturing Key Laboratory, School of Mechanical and Electronic Engineering, Wuhan University of Technology, Wuhan 430070, China

*Corresponding author: **Zhaobing Liu** (School of Mechanical and Electronic Engineering, Wuhan University of Technology, Wuhan 430070, China. Tel.: +86 18641204189, E-mail address: zhaobingliu@whut.edu.cn, ORCID: 0000-0001-8800-0731).

## Abstract

The current apple detection algorithms fail to accurately differentiate obscured apples from pickable ones, thus leading to low accuracy in apple harvesting and a high rate of instances where apples are either mispicked or missed altogether. To address the issues associated with the existing algorithms, this study proposes an improved YOLOv5s-based method, named YOLOv5s-BC, for real-time apple detection, in which a series of modifications have been introduced. Firstly, a coordinate attention (CA) block has been incorporated into the backbone module to construct a new backbone network. Secondly, the original concatenation operation has been replaced with a bidirectional feature pyramid network (BiFPN) in the neck module. Lastly, a new detection head has been added to the head module, enabling the detection of smaller and more distant targets within the field of view of the robot. The proposed YOLOv5s-BC model was compared to several target detection algorithms, including YOLOv5s, YOLOv4, YOLOv3, SSD, Faster R-CNN (ResNet50), and Faster R-CNN (VGG), with significant improvements of 4.6%, 3.6%, 20.48%, 23.22%, 15.27%, and 15.59% in mAP, respectively. The detection accuracy of the proposed model is also greatly enhanced over the original YOLOv5s model. The model boasts an average detection speed of 0.018 seconds per image, and the weight size is only 16.7 Mb with 4.7 Mb smaller than that of YOLOv8s, meeting the real-time requirements for the picking robot. Furthermore, according to the heat map, our proposed model can focus more on and learn the high-level features of the target apples, and recognize the smaller target apples better than the original YOLOv5s model. Then, in other apple orchard tests, the model can detect the pickable apples in real time and correctly, illustrating a decent generalization ability. It is noted that our model can provide technical support for the apple harvesting robot in terms of real-time target detection and harvesting sequence planning.

# 1. Introduction

Apple, as one of the top four fruits in the world, is rich in many vitamins and minerals and has been popular with consumers around the world. According to the statistics of the Food and Agriculture Organization of the United Nations, apples rank second after grapes in global fruit production (Jia et al., 2020). However, most apple fruits are hand-picked, and such production methods are very inefficient. In addition, with an aging population and a large influx of rural labor into the cities, labor costs in the fruit cultivation industry have risen accordingly. All these factors significantly impact the market competitiveness of fruit products. Therefore, it is imperative to harvest apple and other fruits efficiently in real-time and reduce harvesting costs. Fruit harvesting robot based on machine vision can use its information perception to identify and pick fruits. Thus, it can improve efficiency and increase economic benefits, which has become a research hotspot for intelligent agricultural equipment (Lv et al., 2022). However, there are still few products of fruit harvesting robots applied in agriculture, and most of them are relatively low in intelligence and even less in large-scale applications (Zhou et al., 2022 , Fountas et al., 2020). In view of the above situation, it is of great practical significance to study the technology related to fruit harvesting robots.

Within the laboratories, different fruit harvesting robots are studied. Although these fruit-harvesting robots have unique features suitable for specific application scenarios, they all rely on the same core technologies, such as stable mobile platforms, multi-sensor collaboration, advanced machine vision technology, and flexible motion control. Among them, machine vision has been drawing great attention due to the rapid development of artificial intelligence. Over the years, researchers have combined vision technology to recognize and locate fruits to provide technical support for fruit harvesting robots. A comprehensive survey revealed that in the field of machine vision, target detection algorithms have tremendous potential for growth by virtue of their high detection accuracy and easy deployment (Zhao et al., 2016). Note that, the mainstream two-stage target detection algorithms include Faster R-CNN (Ren et al., 2017) and Mask R-CNN (He et al., 2017), while one-stage target detection algorithms are SSD (Liu et al., 2016) and YOLO (You Only Look Once) series including YOLOv3 (Redmon and Farhadi, 2018), YOLOv4 (Bochkovskiy et al., 2020), and YOLOv5 (ultralytics), etc.. Noticeably, the two-stage target detection algorithms generally have higher detection accuracy, but the trained model is large, leading to slow detection speed during practical detection. In contrast, the one-stage target detection algorithm is increasingly used as the preferred solution due to the advantages of the small number of model parameters and rapid detection speed.

The following section will focus on discussion of YOLO, applied to agriculture in the last three years. The YOLO series was pioneered by Redmon and his colleagues and developed on the darknet in versions YOLOv1, YOLOv2, and YOLOv3 (Redmon et al., 2016 , Redmon and Farhadi, 2017 , Redmon and Farhadi, 2018). Numerous iterations have emerged since then, and Bochkovskiy et al. (2020) have continued to build on the darknet and come up with YOLOv4. Unlike previous versions, Ultralytics developed YOLOv5 with the Pytorch framework. YOLOv5 is favored by researchers for its ease of deployment and well detection performance. YOLOv5 has four basic network models: YOLOv5s, YOLOv5m, YOLOv5l, and YOLOv5x. Their feature map depths are progressively deeper, and the model parameters are increased sequentially. Table 1 summarizes the performance of the improved YOLOv5 model in the agricultural domain. In terms of apple detection, Yan et al. (2021) proposed a light target detection method for apple-picking robots based on an improved YOLOv5s algorithm. The bottleneck Cross Stage Partial (CSP) module is redesigned as a bottleneck CSP-2 module. In addition, the squeeze and excitation module in the visual attention mechanism network is inserted into the improved backbone network. The average detection accuracy is 86.75%. Lv et al. (2022) proposed a visual recognition method for detecting apple growth patterns in orchards using the YOLOv5s algorithm. The authors replaced the SiLU activation function in the network with the ACON-C activation function, which improved the accuracy of the algorithm without sacrificing real-time performance. Sun et al. (2022) proposed an improved lightweight apple detection method YOLOv5-PRE for fast apple yield detection in an orchard environment and introduced ShuffleNet and GhostNet lightweight structures in the YOLOv5-PRE model to reduce the size of the model. Xu et al. (2023) proposed an improved YOLOv5 apple grading method. The Mish activation function replaced the original YOLOv5 activation function, and the squeeze excitation module was added to the YOLOv5 backbone. The average accuracy of the improved YOLOv5 algorithm for grading apples under the test set is 90.6%. For the detection of other fruits and vegetables, Yao et al. (2021) developed a YOLOv5-based Kiwifruit defect detection model, called YOLOv5-Ours. The proposed model adds a small target detection layer by embedding SELayer attention to different channels. The average detection accuracy of YOLOv5-Ours reached 94.7%. Wu et al. (2022) constructed a new YOLOv5-B model by enhancing the loss function. Then, the optimal truncation point is obtained by segmenting the contours of the axes utilizing an edge detection algorithm. Experiments show that the average detection accuracy of the proposed model for banana multi-target recognition is 93.2%. Xu et al. (2022) proposed an improved YOLOv5s-based target detection method for Zanthoxylum-picking robots. An improved CBF module is proposed based on the backbone CBH module, and a Specter module is proposed to replace the bottleneck CSP module. Test experiments conducted on NVIDIA Jetson TX2 show that the average inference time is 0.072s. Liang et al. (2023) developed a YOLOv5-SBiC algorithm for late autumn bud recognition. In the algorithm, a transformer module is introduced to speed up the network convergence, an attention mechanism module is used to help the model extract more useful information. Test results show that the proposed algorithm improves the recognition accuracy by 4.0% over the original YOLOv5 algorithm, reaching 79.6%. In the field of pests and diseases detection of fruits and vegetables, Zhang et al. (2022) proposed a new method based on a target detection network, feature extraction, and classifier to detect adjacent wheat ears. The proposed algorithm combines distance-interlinked non-maximum suppression based on the original YOLOv5 to form an improved YOLOv5 target detection network with an average detection accuracy of 90.67% and a detection time of 0.73ms. Qi et al. (2022) implemented the extraction of key features by inserting a squeeze stimulus module into the original YOLOv5 network framework, drawing on the human visual attention mechanism. The model was evaluated on the tomato virus disease test set, and the average detection accuracy was 94.10%. Bao et al. (2023) proposed a DDMA-YOLO-based UAV remote sensing method to detect and monitor tea leaf blight. The algorithm added a multiscale RFB module based on the original YOLOv5 network with

**Table 1**

Performance of the improved YOLOv5 models

| Detection object | Networks model | F1 (%) | mAP (%) | Detection speed (FPS) | GPU | Reference |
|---|---|---|---|---|---|---|
| Apple | Improved YOLOv5s | 87.49 | 86.75 | 66.7 | Nvidia Geforce RTX 2060 | (Yan et al., 2021) |
| Apple | YOLOv5-B | 92.8 | 98.4 | 71 | Nvidia Geforce GTX 1080 | (Lv et al., 2022) |
| Apple | YOLOv5-PRE | 88.88 | 94.03 | 37.04 | Nvidia Quadro P620 | (Sun et al., 2022) |
| Apple | Im-YOLOv5 | 90.74 | 90.6 | 59.63 | Nvidia Geforce GTX 1660Ti | (Xu et al., 2023) |
| Kiwifruit | YOLOv5-Ours | - | 94.7 | 10 | Nvidia GeForce GTX 1050Ti | (Yao et al., 2021) |
| Banana | YOLOv5-B | 94.44 | 93.2 | 111.1 | Nvidia Tesla V100 SXM2 | (Wu et al., 2022) |
| Zanthoxylum | Improved YOLOv5s | - | 94.5 | 88.33 | Nvidia GeForce RTX 3060 Laptop | (Xu et al., 2022) |
| Shoots of litchi | YOLOv5-SBiC | - | 79.56 | 55.6 | Nvidia GeForce RTX 3090 | (Liang et al., 2023) |
| Fusarium head blight in wheat | YOLOv5-DIOU | 87.95 | 91.18 | - | Nvidia GeForce RTX 3060 | (Zhang et al., 2022) |
| Tomato virus disease | SE-YOLOv5 | 89.39 | 94.1 | 50.63 | Nvidia GeForce RTX 2060 Super | (Qi et al., 2022) |
| Tea leaf blight | DDMA-YOLO | 71.6 | 76.8 | - | Nvidia GeForce RTX 2060 | (Bao et al., 2023) |
| Passion fruit pests | Improved YOLOv5 | 95.54 | 96.51 | 129.87 | - | (Li et al., 2023) |

dual-dimensional mixed attention (DDMA) in the neck, and the average detection accuracy was 76.8%. Li et al. (2023) developed a fast and lightweight improved YOLOv5 detection algorithm. Based on the original YOLOv5 model, a new point-line distance loss function is presented. In addition, an attention module is added to the network for adaptive attention, which can attend to the target object passion fruit pests in both channel and spatial dimensions. The average detection accuracy is 96.51%.

Considering the above-mentioned discussions, real-time apple detection methods with lightweight models need to be further developed. Through a comprehensive survey of the improved YOLO target detection methods in the agricultural field, although most of the existing detection models have relatively high recognition accuracy, their increased complexity, parameters, and hardware requirements usually lead to low real-time performance. Therefore, designing a lightweight real-time apple detection algorithm is necessary to meet the requirements of real-time recognition of picking robots while ensuring recognition accuracy. In this paper, we propose an improved YOLOv5s-based real-time apple detection method to overcome the limitations of current apple recognition techniques. First, a CA block has been incorporated into the backbone module to construct a new backbone network. Then, the original concatenation operation has been replaced with a BiFPN in the neck module. Additionally, a novel detection head is added to the head module to spot smaller and far-off targets within the field of view. It is worth noting that the proposed model can increase recognition accuracy while ensuring high recognition speed.

## 2. Data acquisition and preprocessing

### *2.1 Apple images acquisition*

In this research, the dataset was from the Agricultural Automation and Robotics Laboratory at Washington State University that was originally utilized to estimate yields in robotic harvesting (Lu and Young, 2020). To obtain the dataset, the lab researchers installed the image sensor behind the prismatic gantry of the robot. The distance between the sensor and the tree was nearly 1.5 meters. Fig. 1 displays the apple images in the dataset from early morning to dusk. In this work, we took 1750 apple RGB images from the original dataset as the new dataset and initially divided this dataset according to the ratio of 0.85:0.15, where 1487 images were in the training set and 263 images were in the test set. There was no overlapping between the two sets.

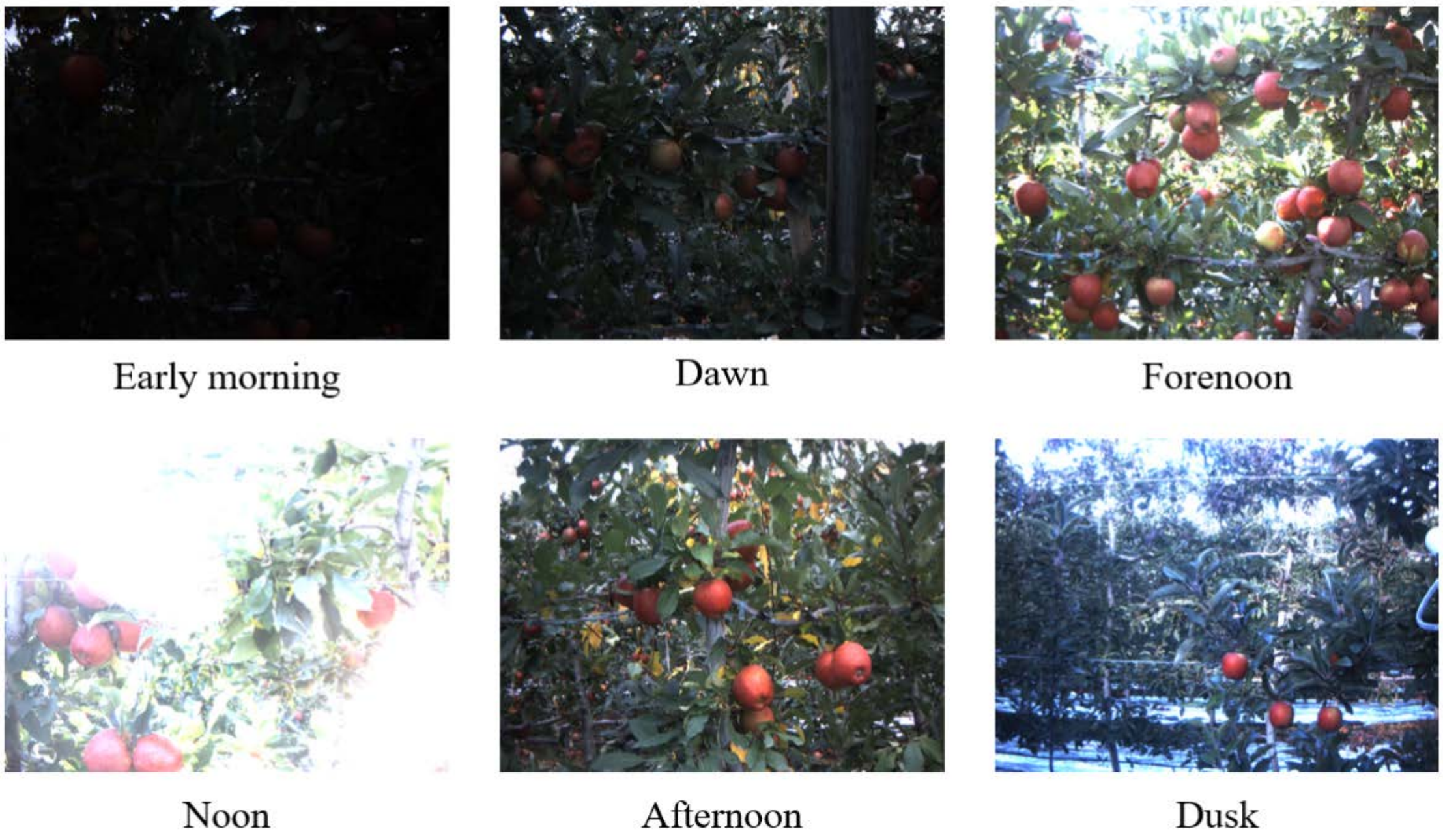

**Fig 1.** Apple images from various points in time (Lu and Young, 2020)

## *2.2 Image labelling*

The labelling software (Labelimg) is utilized to classify and label apple images that are visible to the human eyes after acquisition, as shown in Fig. 2. Due to the intricate environment in apple orchards, the apple images are separated into two categories: graspable and ungraspable, with the corresponding labels ‘apple’ and ‘block’. In detail, apples are categorized based on the following criteria:

(i) Classification 1: Apples that are obstructed by leaves and branches are categorized as ungraspable apples.

(ii) Classification 2: Small target apples that can be observed despite being far away are categorized as graspable apples, which provides valuable data for training models for small targets.

(iii) Classification 3: Large target apples that are close enough and can be observed directly are categorized as graspable apples.

(iv) Classification 4: Apples that are recognized in both bright daylight and insufficient light are categorized as graspable. This adds complexity to the data and enhances the robustness of the model.

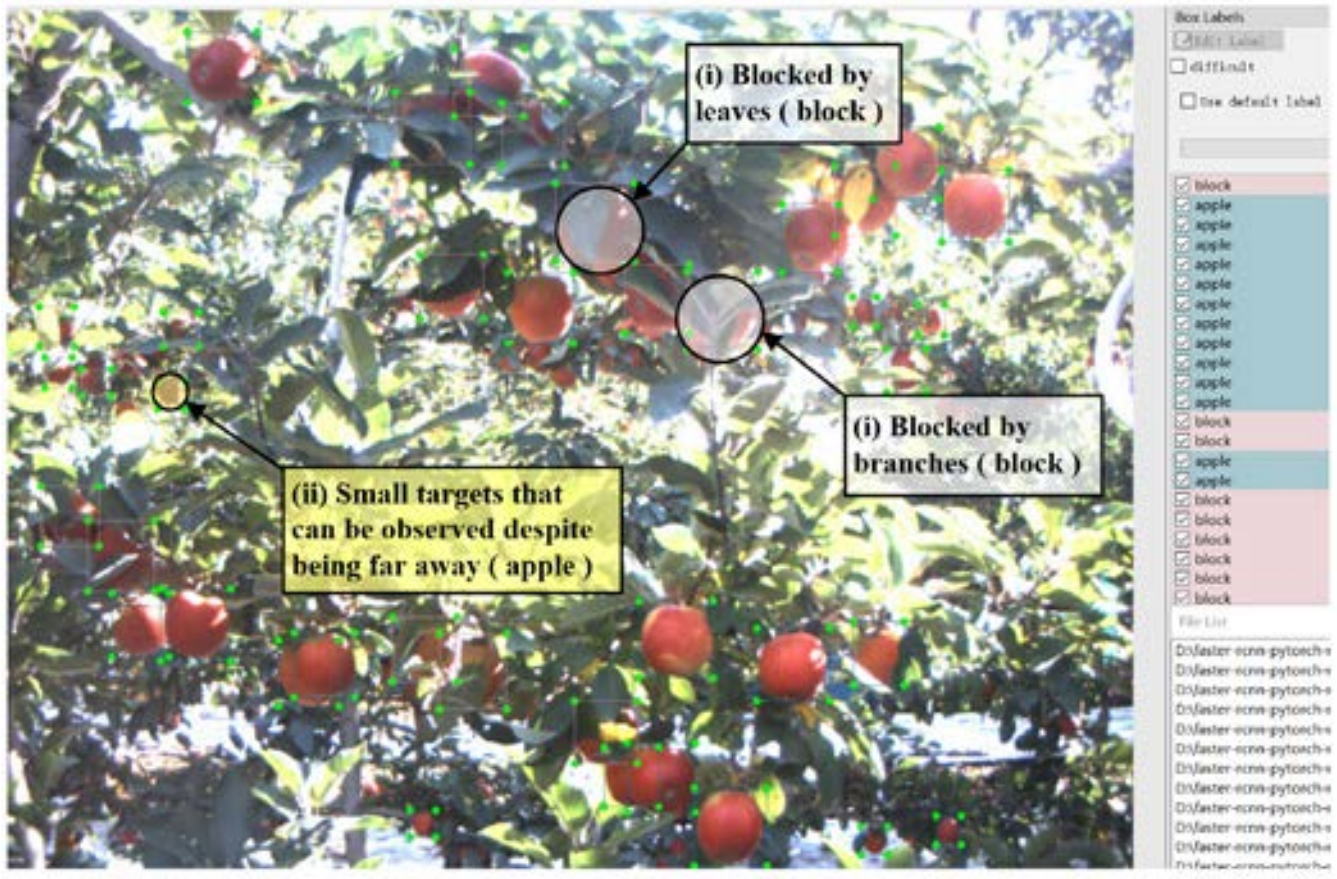

(a) Classification 1 and Classification 2

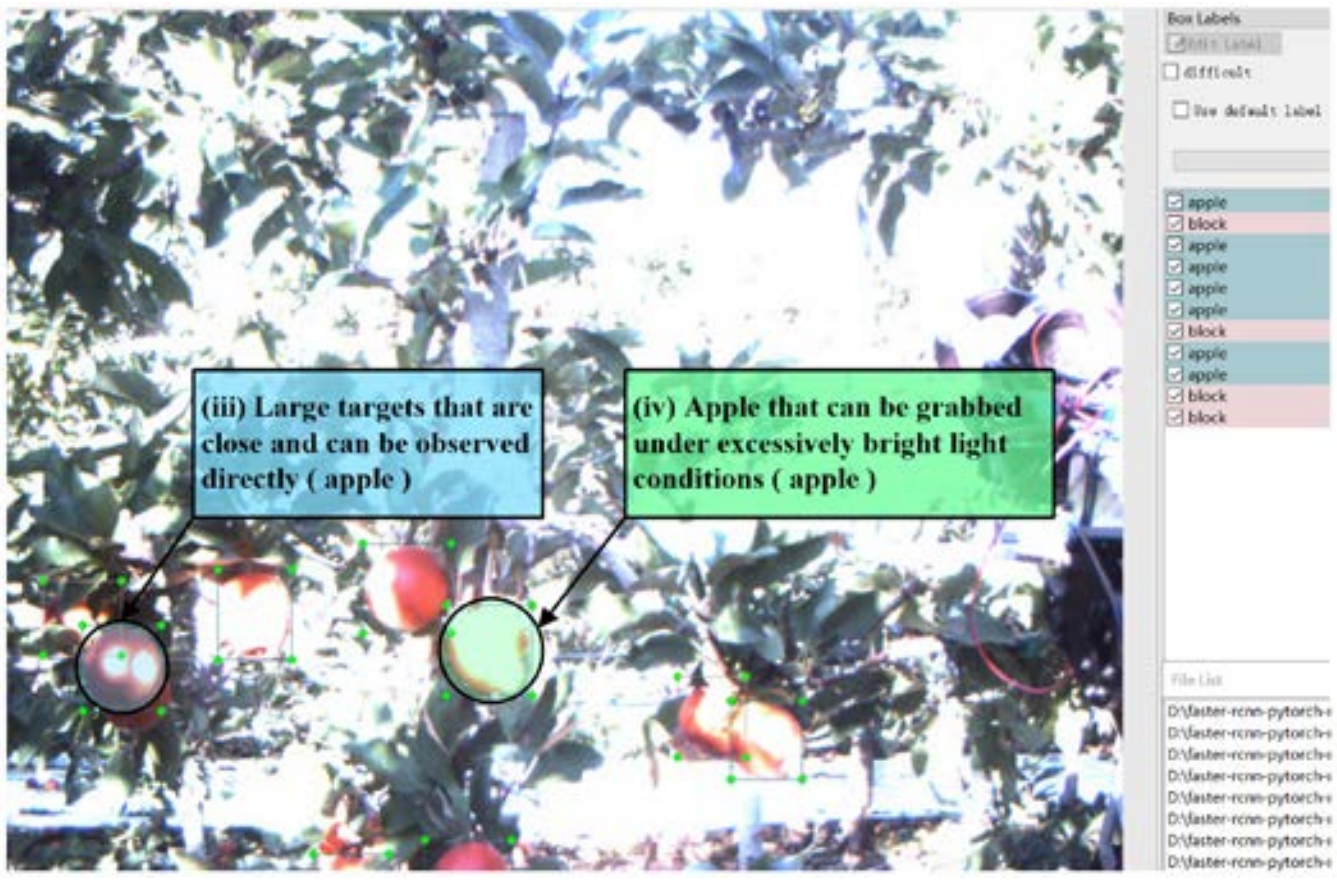

(b) Classification 3 and Classification 4

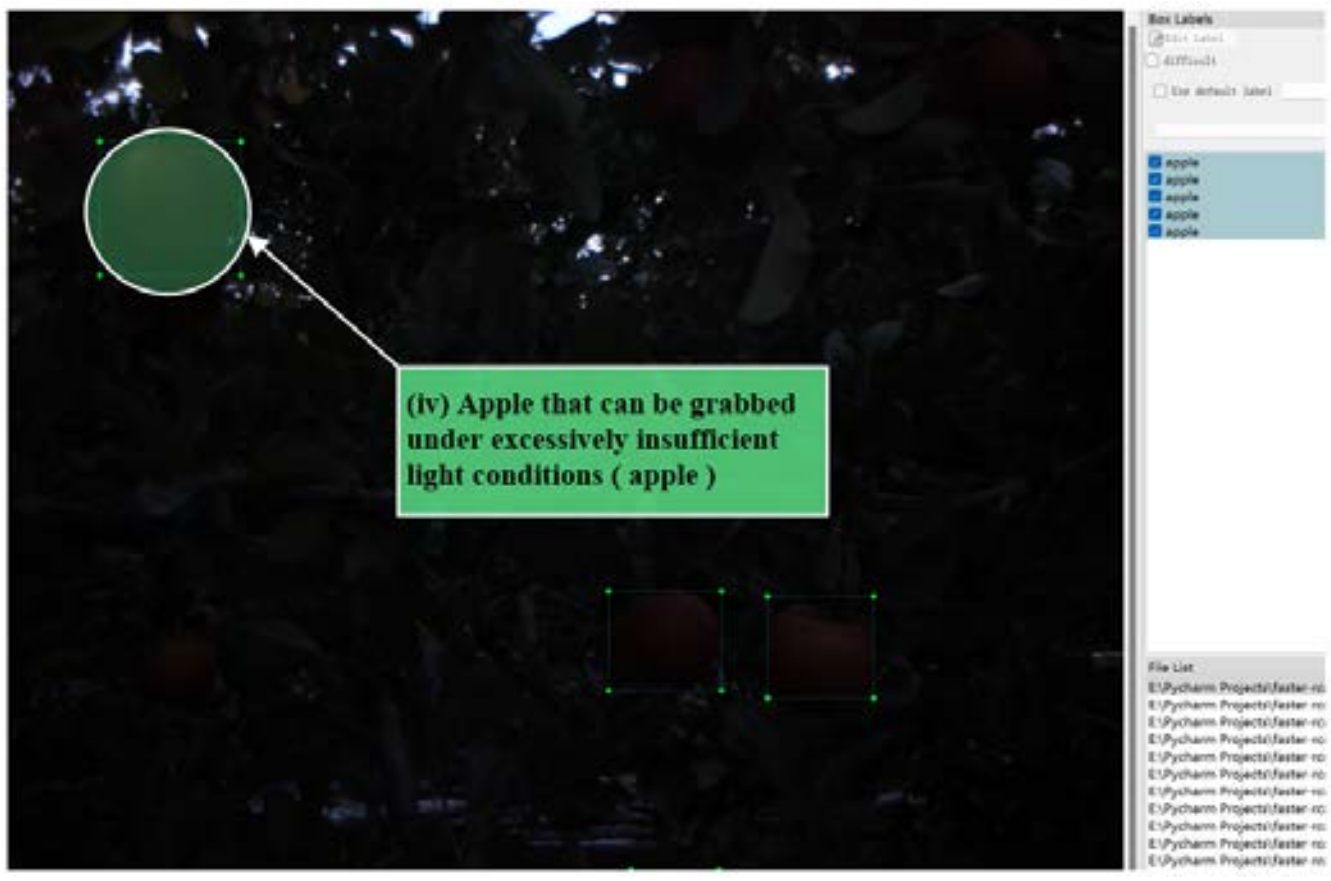

(c) Classification 4

**Fig. 2.** Apple image labelling in the software

## *2.3 Image augmentation*

The quality of the training set plays a pivotal role in determining the capability of the convolutional neural network (CNN) model to identify apples accurately. If the training set is too small, it can lead to overfitting of the model, which may impede its performance in new or unknown environments. Image augmentation involves enhancing the visual quality of an image and augmenting its specific features by applying a series of processes. This methodology can effectively enlarge the size and diversity of the training set and improve the generalization ability of the CNN model. Specific image enhancement methods were selected based on application scenarios and data characteristics. We have selected eight data augmentation methods based on our own scenario requirements. These methods included random contrast, edge enhancement, contrast-limited adaptive histogram equalization (Clahe), motion blur, perspective transformation, adding salt and pepper noise, max pool, and changing color temperature (Yan et al., 2021). A total of 11896 enhanced images were generated by these methods from the initial training set of 1487 images, so the new training set consists of 13383 images. Fig. 3 illustrates the eight different image augmentation methods that have been used on each image.

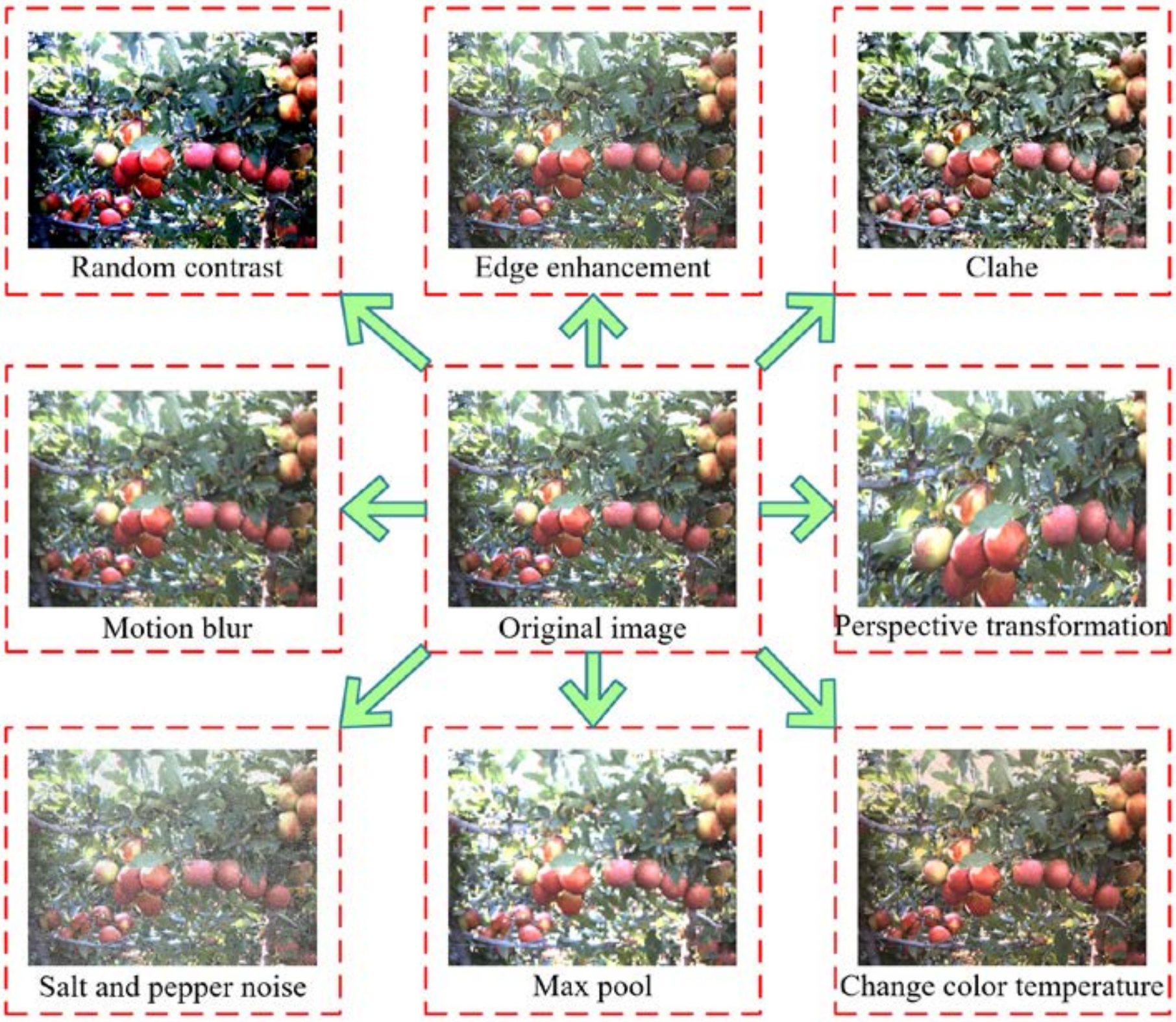

**Fig 3.** Enhanced images using different image enhancement methods

# 3. Methods

## *3.1 YOLOv5s*

The YOLOv5 algorithm is a one-stage target detection algorithm that generates class probabilities and position coordinate values for objects without requiring region proposals. It is one of the most popular target detection algorithms among agricultural researchers and its network structure can be divided into four modules: input, backbone, neck, and head. The input module uses mosaic data augmentation, adaptive anchor frame calculation, and adaptive image scaling operations. The backbone network consists of focus and Cross Stage Partial (CSP) structures. The neck module utilizes a Feature Pyramid Network (FPN) and Path Aggregation Network (PAN) structure. The CIoU loss function is used as the loss function of the bounding box in the head module. Readers can refer to the official code (https://github.com/ultralytics/YOLOv5) for more details. Because the YOLOv5s model has the fewest parameters among the four models officially provided by YOLOv5, it is in line with the trend of lightweight and easier to deploy on fruit harvesting robots, thereby satisfying the effect of real-time grasping. Therefore, in this study, we choose it as the research candidate.

## *3.2 CA block*

The attention mechanism is essential in identifying targets as it enables the model to concentrate on crucial parts of images, thereby improving accuracy and efficiency in detection. Hou et al. (2021) proposed a CA mechanism that integrates position information into channel attention. In more detail, CA decomposes channel attention into two one-dimensional feature encoding processes that aggregate features along two spatial directions, respectively. This allows for capturing remote dependencies in one spatial direction while maintaining accurate location information in the other spatial direction. The resulting feature maps are then encoded as a pair of direction-aware and position-sensitive attention maps, respectively, which can be applied complementarily to the input feature maps to enhance the representation of the object of interest. In addition, CA has the property of portability and can be flexibly embedded into CNN. Considering collectively, we choose it as our attention mechanism component to be introduced into the YOLOv5 network in this experiment. The specific operation of CA is divided into 2 steps: coordinate information embedding and CA generation. Fig. 4 shows the structure of the CA block.

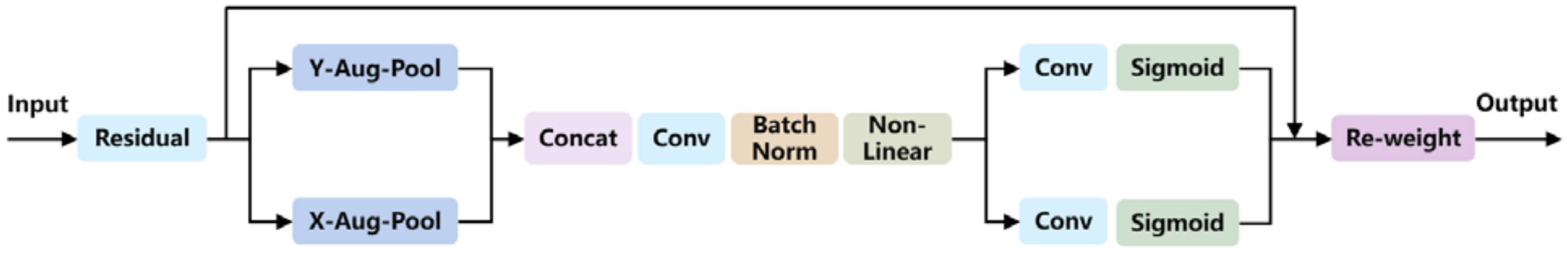

**Fig 4.** Structure of the CA block. The terms 'X-Aug-Pool' and 'Y-Aug-Pool' refer to the one-dimensional horizontal and vertical global pools, respectively.

### *3.2.1 Coordinate information embedding*

The CA block is designed to obtain attention to the width and height of the image and encode the exact position information. First, the input feature map $X$ is divided into two directions, height $h$, and width $w$, and is pooled globally to obtain the feature maps in both directions. Thus, the output of the $c$-th channel with height $h$ can be expressed as:

$$z_c^h(h) = \frac{1}{W} \sum_{0 \le i < W} x_c(h,i) \tag{1}$$

Similarly, the output of the $c$-th channel with width $w$ can be written as:

$$z_c^w(w) = \frac{1}{H} \sum_{0 \le j < H} x_c(j, w) \tag{2}$$

### 3.2.2 CA generation

After the transformation in information embedding, the obtained feature maps in two directions are concatenated together. Then, a $1\times1$ convolution kernel is used to convolve the concatenated feature maps. Further, the batch normalization of the convolved feature maps is performed to obtain the feature map $F_1$, and the non-linear activation function is used to activate the feature map $F_1$ to obtain the feature map $f$. The above process can be expressed as follows:

$$f = \delta\left(F_1\left(\left[z^h, z^w\right]\right)\right) \tag{3}$$

Then, $f$ is sliced into two separate tensors $f^h$ and $f^w$ along the spatial dimension, and next, the feature maps $f^h$ and $f^w$ are transformed to the same number of channels as the input feature map $X$ using two $1\times1$ convolutions $F_h$ and $F_w$. The sigmoid activation function is used to activate it. The equation is expressed as follows:

$$g^h = \sigma\left(F_h\left(f^h\right)\right) \tag{4}$$

$$g^w = \sigma\left(F_w\left(f^w\right)\right) \tag{5}$$

Finally, the output of CA block $Y$ can be written as:

$$Y = X \times g^h \times g^w \tag{6}$$

## 3.3 BiFPN

To enhance the efficiency of the model, Tan et al. (2020) developed a weighted bidirectional feature pyramid network. The main objective of this structure is to create a bidirectional connectivity mechanism based on the FPN, which allows information to flow in both directions and gradients to propagate throughout the network. This network is capable of multi-scale feature fusion of feature maps from different resolutions, thus improving the overall performance of the network. Since it is a versatile network, BiFPN can be seamlessly integrated with different neural network architectures, thus enhancing the generalization capability and stability of the network, for a wide range of image segmentation tasks. Fig. 5 meticulously illustrates the comparison of FPN and BiFPN structures, where $P_i$ represents a feature level with resolution of $1/2^i$ of the input images.

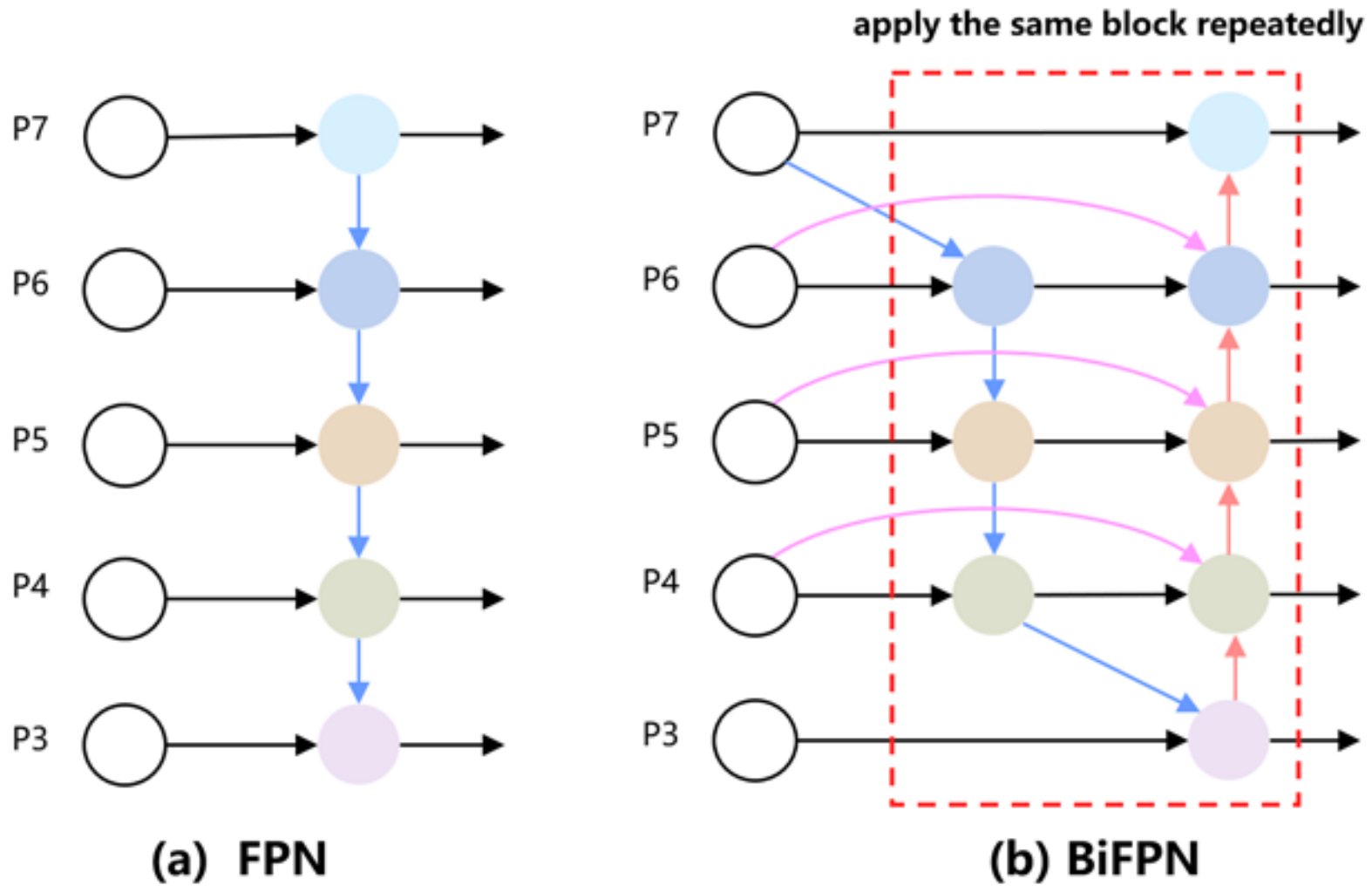

**Fig 5.** Comparison of FPN and BiFPN. For P3-P7, it is multi-scale features from level 3 to level 7.

### *3.4 YOLOv5s-BC*

In actual detection, the YOLOv5s algorithm can detect apples with high recognition. However, due to the interference and influence of the complex environment in the orchard, small target apples that are far away are usually ignored by the algorithm. Considering that the obscured apples are mistaken as targets, this leads to the robot that is not able to grab the apples by estimating their position and posture correctly. Consequently, we propose an improved YOLOv5s algorithm, named YOLOv5s-BC, by making several modifications as follows.

(i) The CA block is implemented in both the backbone and neck modules. Integration of the CA mechanism in the backbone enhances the feature pyramid representation capability and the model's abstraction capability. With the inclusion of the CA mechanism in the neck, there exist potential enhancements in the differentiating ability of the feature pyramid, aiding the model to differentiate targets at varying scales, and improving the detection ability of smaller targets.

(ii) The neck module introduces a BiFPN that receives feature maps from the backbone at various scales. It constructs a more stable and precise feature pyramid by fusing features at multiple levels and adjusting adaptive weights. Then, the neck module sends this feature pyramid to the subsequent detection head for detecting targets.

(iii) A new detection head is incorporated into the head module to enhance target detection performance (Zhu et al., 2021). The addition of this head enables the detection of smaller targets situated farther away by utilizing high-resolution feature maps, which results in higher accuracy in both target detection and localization.

Fig. 6 depicts the network architecture of the enhanced algorithm in detail.

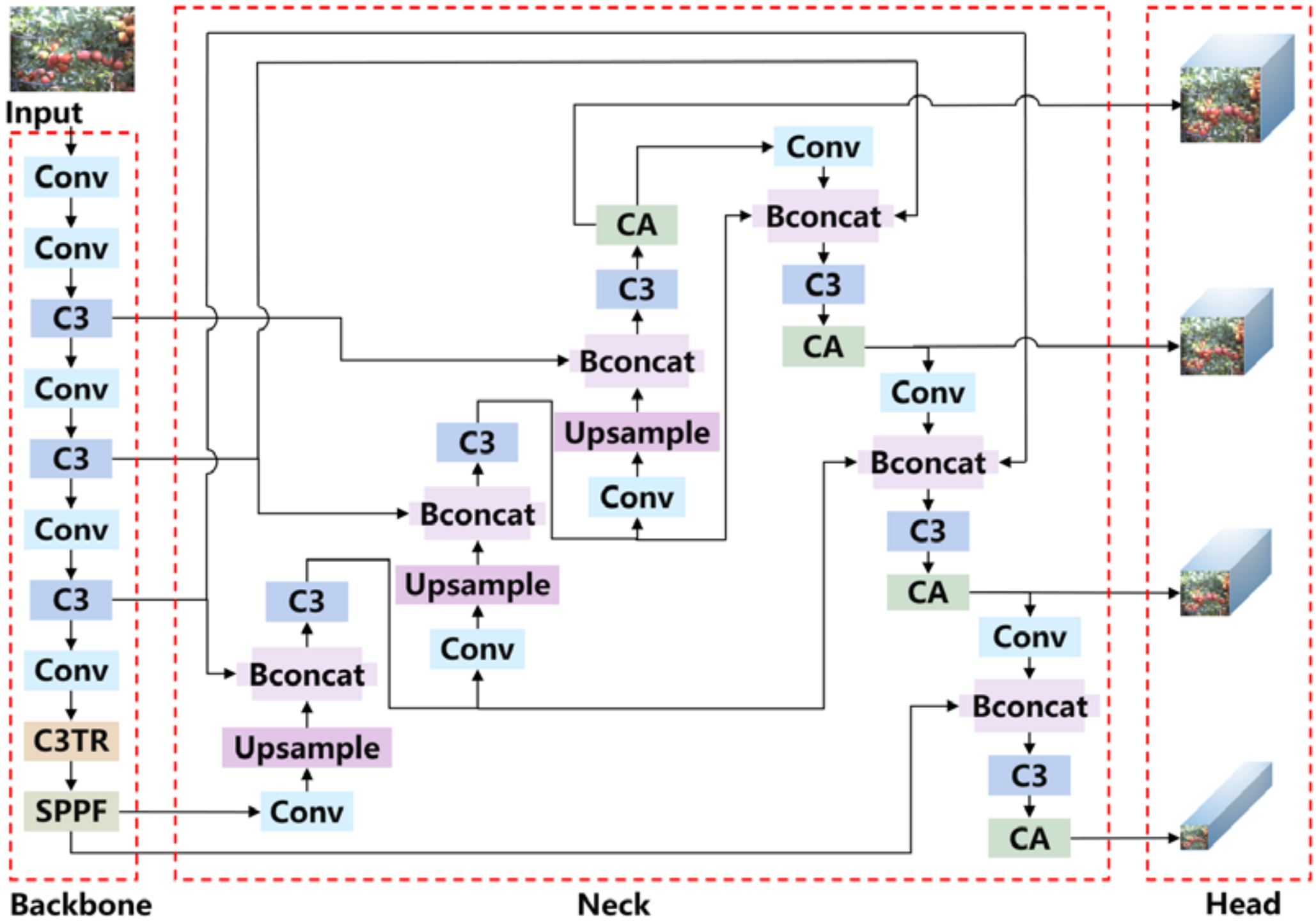

**Fig 6.** The network architecture of YOLOv5s-BC

## *3.5 Model evaluation index*

To evaluate the performance of the established model, several indexes are discussed in this section. True positives (TP) refer to positive samples that are categorized correctly. True negatives (TN) are negative samples that are identified accurately. False positives (FP) are negative samples that are mislabeled as positives. False negatives (FN) occur when positive samples are wrongly labeled as negative. *Precision* and *Recall* are defined in Eq. (7) and Eq. (8), respectively. The calculation formula for *Accuracy* is shown in Eq. (9). $F1$ Score has become a metric often used in statistics to measure the accuracy of classification models due to the fact that it combines both the precision and recall of a classification model. It can be calculated by Eq. (10).

$$Precision = \frac{TP}{TP + FP} \tag{7}$$

$$Recall = \frac{TP}{TP + FN} \tag{8}$$

$$Accuracy = \frac{TP + TN}{TP + FP + FN + TN} \tag{9}$$

$$F1 = \frac{2 \times Precision \times Recall}{Precision + Recall} \tag{10}$$

The precision-recall (P-R) curve is plotted with the horizontal coordinate as the recall rate $R$ and the vertical coordinate as the precision rate $P$. The area enclosed by this curve is the average precision (AP). The calculation of $AP$ is based on the Eq. (11).

$$AP = \int_0^1 P_{(R)} dR \tag{11}$$

Mean average precision (mAP) is another commonly used evaluation metric in target detection models, which is the average of AP rate of each category.

The Intersection over Union (IoU) ratio is a crucial concept in target detection. It is the ratio of the intersection areas and union areas between the predicted box and the ground truth box, with non-deformation and non-negativity on the scale, as shown in Eq. (12). $B^{gt}$ is the area of ground truth box, and $B$ is the area of predicted box. $w^{gt}$ and $h^{gt}$ are the width and height of the ground truth box respectively. Similarly, $w$ and $h$ are the width and height of the predicted box respectively (see Fig. 7(a)).

$$IoU = \frac{\left|B \cap B^{gt}\right|}{\left|B \cup B^{gt}\right|} \tag{12}$$

In Fig. 7(a), it can be observed that two boxes did not intersect and as a result, IoU value is equal to zero which is insufficient to indicate their mutual distance. Additionally, when the loss is equivalent to zero, there is no backward transfer of gradient, and therefore, no learning progress can occur. Consequently, IoU falls short in providing a robust representation of their intersection. Fig. 7(b) demonstrates that in both cases the IoU remains equal, however, their degree of overlap is different. To address this, numerous solutions have been proposed recently to enhance the IoU calculation. In this research, we adopt the original YOLOv5s selection, namely Complete IoU (CIoU), to compute the box loss (Box Loss).

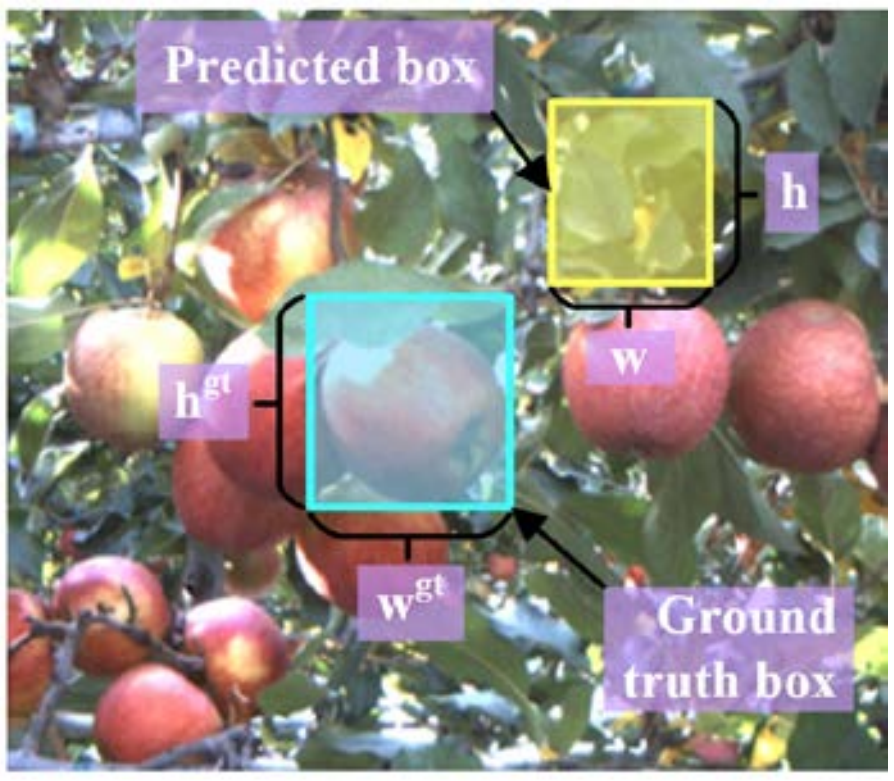

(a) IoU=0

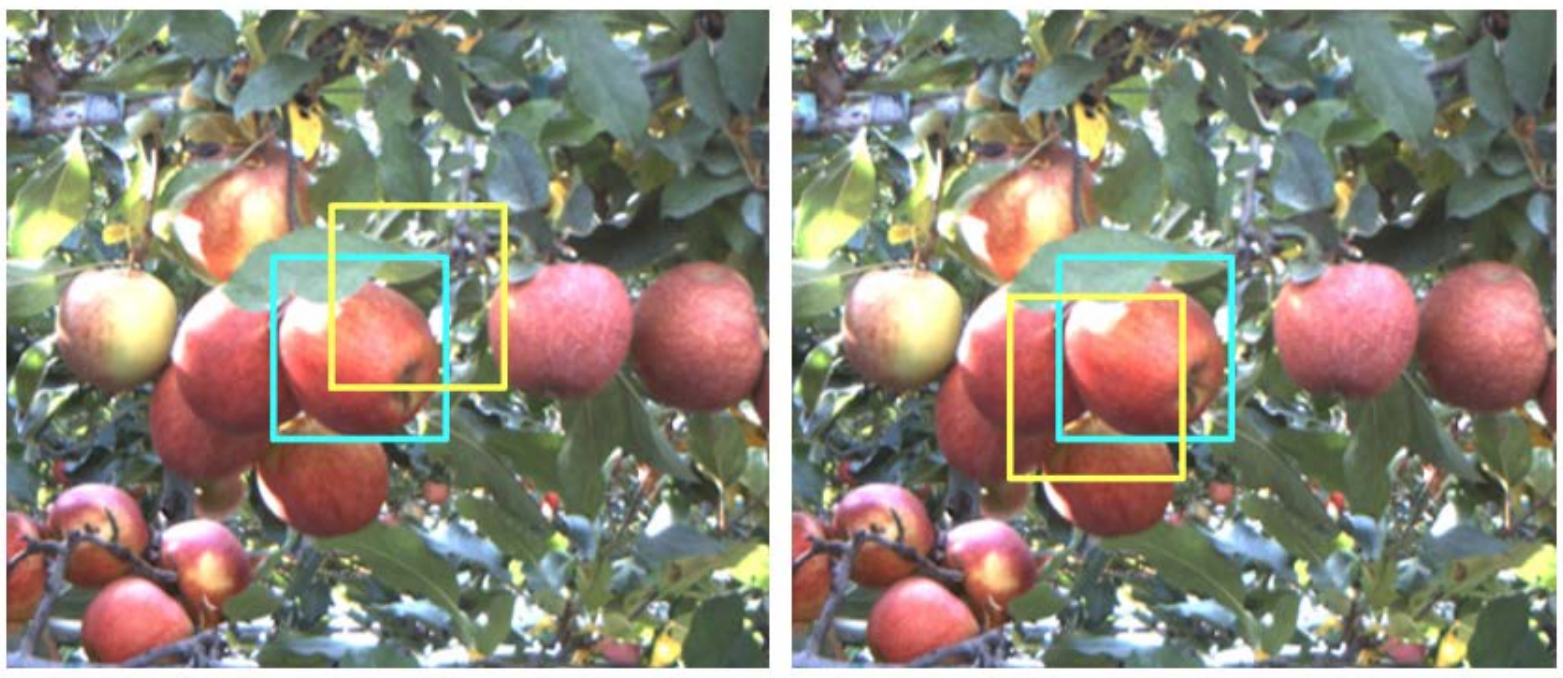

(b) IoU=0.66

Fig 7. IoU values for different situations

The CIoU loss is proposed considering that the consistency of the bounding box aspect ratio is an important geometric factor (Zheng et al., 2020). $\alpha$ is a positive trade-off parameter (See Eq. (13)), $\upsilon$ is the similarity of the metric aspect ratio (see Eq. (14)), where $b$ and $b^{gt}$ denote the central points of ground truth box and predicted box, $distance(\cdot)$ is the Euclidean distance, and $length(\cdot)$ is the diagonal length of the smallest closed box that covers both boxes. CIoU loss is calculated by Eq. (15).

$$\alpha = \frac{\upsilon}{(1 - IoU) + \upsilon} \tag{13}$$

$$\upsilon = \frac{4}{\pi^2}\left(arctan\frac{w^{gt}}{h^{gt}} - arctan\frac{w}{h}\right)^2 \tag{14}$$

$$Loss_{CIoU} = 1 - CIoU = 1 - IoU + \left(\frac{distance(b, b^{gt})}{length(B, B^{gt})} + \alpha\upsilon\right) \tag{15}$$

The Binary Cross Entropy (BCE) loss function is utilized to compute both classification loss (Cls Loss) and object loss (Obj Loss), as demonstrated in Eq. (16). $n$ is the total number of samples, $y_i$ is the category which the $i$ sample belongs to, and $x_i$ is the predicted probability of the $i$ sample.

$$Loss_{BCE} = -\frac{1}{n}\sum_{i=1}^{n}\left[y_i ln(x_i) + (1 - y_i) ln(1 - x_i)\right] \tag{16}$$

The loss function measures the difference between predicted and ground truth information. A lower loss function value indicates greater similarity between the predicted and ground truth information. So the loss function is also an essential index to evaluate the target detection model. The loss function for our model is divided into three major parts: Box Loss, Cls Loss, and Obj Loss. It is the weighted sum of these losses (See Eq. (17)).

$$Loss = coef_{box} \times Loss_{box} + coef_{cls} \times Loss_{cls} + coef_{obj} \times Loss_{obj} \tag{17}$$

## 4. Experiments and discussion

### *4.1. Experimental setup*

In this experiment, the training and testing of the model were done on the server. The server configuration parameters are shown in Table 2. In addition, the YOLOv5s-BC network employs stochastic gradient descent (SGD) as the optimizer, with specific hyper-parameters shown in Table 3. Additionally, the model was trained for 200 epochs, and the batch size for model training was set to 16. The input image had a size of 640 pixels by default. After the above training parameters are determined, the model can be trained accordingly.

**Table 2**
**Server configuration parameters**

| Parameters | On the server |
|---|---|
| **Operating system** | Ubuntu 18.04 |
| **GPU** | RTX A4000 (16GB) |
| **CPU** | Intel Xeon Gold 5320 |
| **Deep learning framework** | Pytorch 1.8.1 |
| **Programming Language** | Python 3.8 |

**Table 3**
**Hyper-parameters**

| Hyper-Parameters | Value |
|---|---|
| **Initial value of learning rate** | 0.01 |
| **Momentum** | 0.937 |
| **Weight decay** | 0.0005 |
| **Box loss coefficient** | 0.05 |
| **Cls loss coefficient** | 0.5 |
| **Obj loss coefficient** | 1.0 |

## *4.2 Experimental results*

### *4.2.1 Comparison of different target detection algorithms*

Fig. 8 and Fig. 9 display the mAP and loss values of the YOLOv5s model with different strategies. The first strategy (strategy 1) involves adding the CA block to the network which corresponds to the YOLOv5s-CA curve in the figure. The second strategy (strategy 2) incorporates the idea of BiFPN, which is represented as the YOLOv5s-BiFPN curve in the figure. The third strategy (strategy 3) combines both strategy 1 and strategy 2, and further adds a new detection head, which is illustrated as the YOLOv5s-BC curve in the same figure. Note that, integrating either strategy 1 or strategy 2 solely into the YOLOv5s network achieves better performance compared to the original YOLOv5s algorithm. Our proposed YOLOv5s-BC model takes advantage of both strategy 1 and strategy 2, and the test results show that it has a faster convergence speed and higher detection accuracy than other models. This further demonstrates the flexibility and transferability of these strategies and the superior performance of fusing diverse strategies.

The efficacy of the YOLOv5s-BC was further tested by evaluating its performance against several other prominent target detection models, namely YOLOv8, YOLOv4, YOLOv3, SSD, Faster R-CNN (VGG), and Faster R-CNN (ResNet50). Specifically, Faster R-CNN (VGG) and Faster R-CNN (ResNet50) employ VGG and ResNet50 as their respective backbone networks. All eight models were trained using the same training dataset and parameters determined previously. The training results of different target detection algorithms are presented in Table 4. It shows that the improved YOLOv5s-BC model achieves the 88.7% mAP on the test sets, outperforming the original YOLOv5s, YOLOv4, YOLOv3, SSD, Faster R-CNN (ResNet50), and Faster R-CNN (VGG) models by 4.6%, 3.6%, 20.48%, 23.22%, 15.27%, and 15.59%, respectively.

On the other hand, the YOLOv5s-BC model demonstrates 37.57% decrease in detection speed in comparison to the original YOLOv5s. Nevertheless, it provides a significant improvement over YOLOv4, YOLOv3, SSD, Faster R-CNN (ResNet50), and Faster R-CNN (VGG), by 434%, 331%, 240%, 201%, and 72%, respectively. This observation highlights the superiority of detection speed of the one-stage target detection algorithm over the two-stage target detection algorithm. It is essential to note that models designed for mobile devices with limited resources require lightweight. Therefore, the number of model parameters is an important index that assesses the model performance. The number of network layers and model parameters are increased in our proposed model due to the CA blocks embedded and a new detection head added. It is noted that the weight file of YOLOv5s-BC is 21.9% bigger than the original YOLOv5s. However, it is smaller than YOLOv8, YOLOv4, YOLOv3, SSD, Faster R-CNN (ResNet), and Faster R-CNN (VGG) by 4.7, 239.3, 229.6, 74.4, 96.8, and 530.2 Mb, respectively. In conclusion, although our proposed method is slightly inferior to the original YOLOv5s model in terms of detection speed and the number of model parameters, it is higher than the original YOLOv5s model in terms of detection accuracy. The overall performance of our proposed model is also the highest when compared with other target detection algorithms.

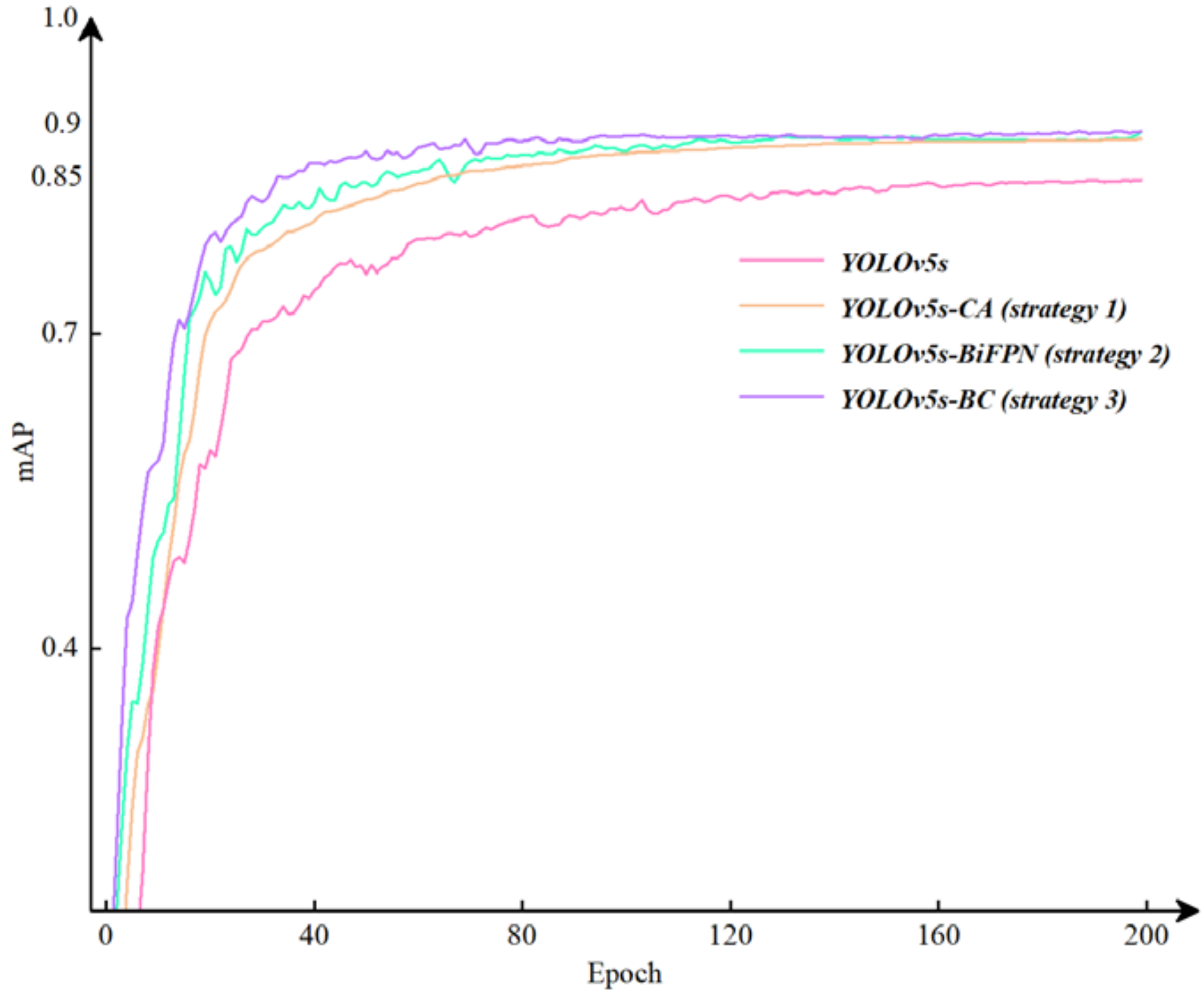

**Fig 8.** The mAP values of the YOLOv5s model with different improvement strategies

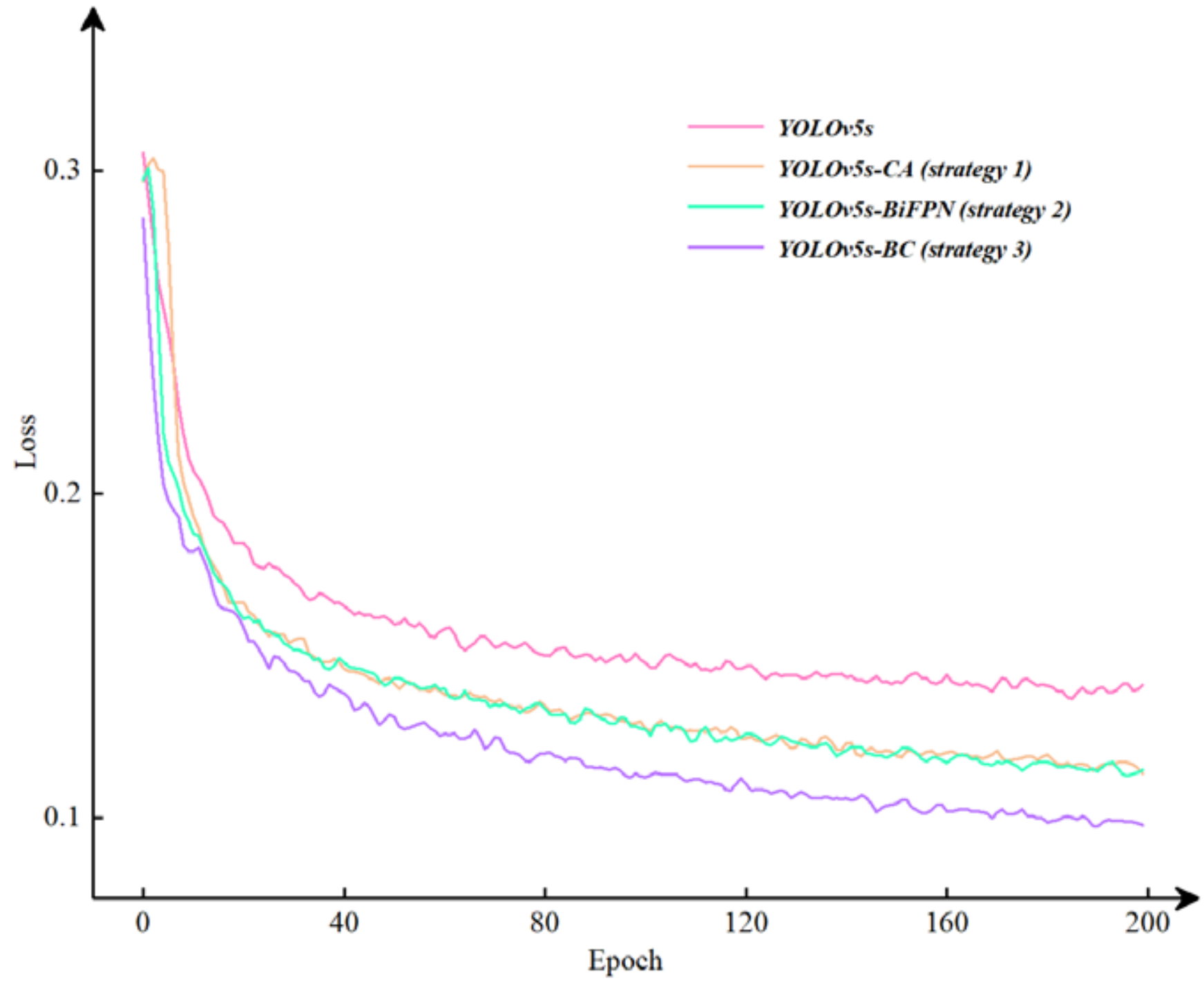

**Fig 9.** The loss values of the YOLOv5s model with different improvement strategies

**Table 4**
**Comparison of detection results of different models**

| **Models** | mAP (%) | AP(Block) (%) | F1 (%) | Detection speed (FPS) | Weight size (Mb) |
|---|---|---|---|---|---|
| **Faster R-CNN (VGG)** | 76.74 | 63.18 | 70.50 | 32.09 | 546.9 |
| **Faster R-CNN (ResNet50)** | 76.95 | 62.44 | 70.50 | 18.34 | 113.5 |
| **SSD** | 68.1 | 50.86 | 68.12 | 16.27 | 91.1 |
| **YOLOv3** | 73.62 | 58.72 | 68.88 | 12.81 | 246.3 |
| **YOLOv4** | 85.62 | 76.11 | 81.31 | 10.35 | 256.0 |
| **YOLOv5s** | 84.8 | 74.2 | 79.83 | 88.5 | 13.7 |
| **YOLOv8s** | 88.6 | 80.4 | 84.0 | 56.49 | 21.4 |
| **YOLOv5s-BC** | 88.7 | 80.5 | 84.32 | 55.25 | 16.7 |

The P-R curves in Fig. 10 depict the performance of the proposed model by comparing its prediction results with the true labels at different thresholds. A model is considered to perform better when its P-R curves for different categories of targets are closer to the upper right corner. Specifically, Table 5 displays the P-R curve and F1 values of the proposed model, while Table 6 illustrates the accuracy of the proposed model when performed on the test set. For the pickable apples category, F1 is 91.6%. For the non-pickable apples category, F1 is 77.0%. This is due to the presence of leaves or overlapping apples obscuring them, making it difficult for the model to learn the complex high-level features. The overall F1 reaches 84.32%, which is the highest score among these algorithms. The detection accuracy reaches 99.8% for the class of graspable apples and 98.55% for the class of ungraspable apples. It is indicated that our model does not overfit on the test set and can detect new apple images well. Additionally, the model achieves a detection speed of over 55 FPS during video detection, demonstrating excellent recognition accuracy and efficiency in the real-time detection. Therefore, our model meets the standard requirements for mobile deployment.

.

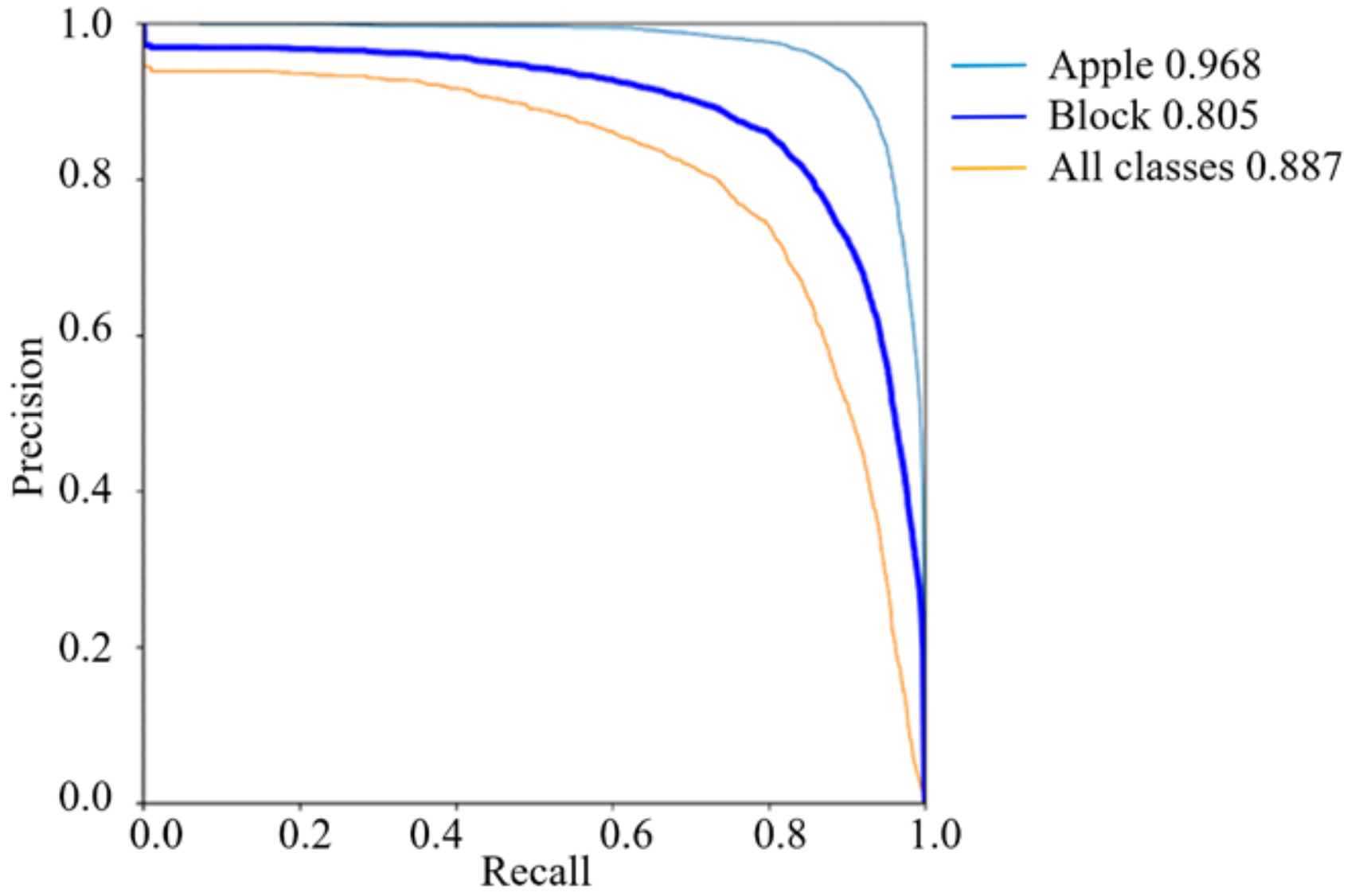

**Fig 10.** The P-R curves of the proposed YOLOv5s-BC model

**Table 5**
**Results of the proposed model**

| Category | Apple (%) | Block (%) | Mean (%) |
|---|---|---|---|
| **Precision** | 91.3 | 74.7 | 83.0 |
| **Recall** | 91.9 | 79.5 | 85.7 |
| **F1** | 91.6 | 77.0 | 84.3 |

**Table 6**
**Accuracy of the proposed model**

| Category | Apple | Block |
|---|---|---|
| **Ground truth** | 6259 | 5101 |
| **Detection results** | 6249 | 5027 |
| **Accuracy** | 99.8% | 98.55% |

### *4.2.2 Further test of the apple detection model*

After conducting the aforementioned experiments, we have determined that the YOLOv5s-BC model offers the most optimal comprehensive performance, which satisfies the prerequisites for detecting apples in real time.

To further assess its accuracy in identifying the morphological attributes of apples, the feature maps of the detection layers were exhibited as heat maps. We chose picture number 1328 from the test set as the display image for conducting comparison experiments of YOLOv5s-BC and YOLOv5s models. Fig. 11 illustrates the heat maps of both YOLOv5s and YOLOv5s-BC at the minimum detection layer. The YOLOv5s-BC model includes a new prediction head to enhance recognition of smaller objects that may be concealed by leaves or located far away. Detection results at small and medium scales reveal that the YOLOv5s model only provides a rough indication of the target location, which includes unnecessary information like leaves and branches. In comparison, our proposed model can identify the target more accurately while avoiding incorporating irrelevant details like leaves and branches, especially on small and medium scales. This is due to the CA mechanism, which enables the model to better focus on the most relevant parts of the image and thus enhances the overall detection accuracy.

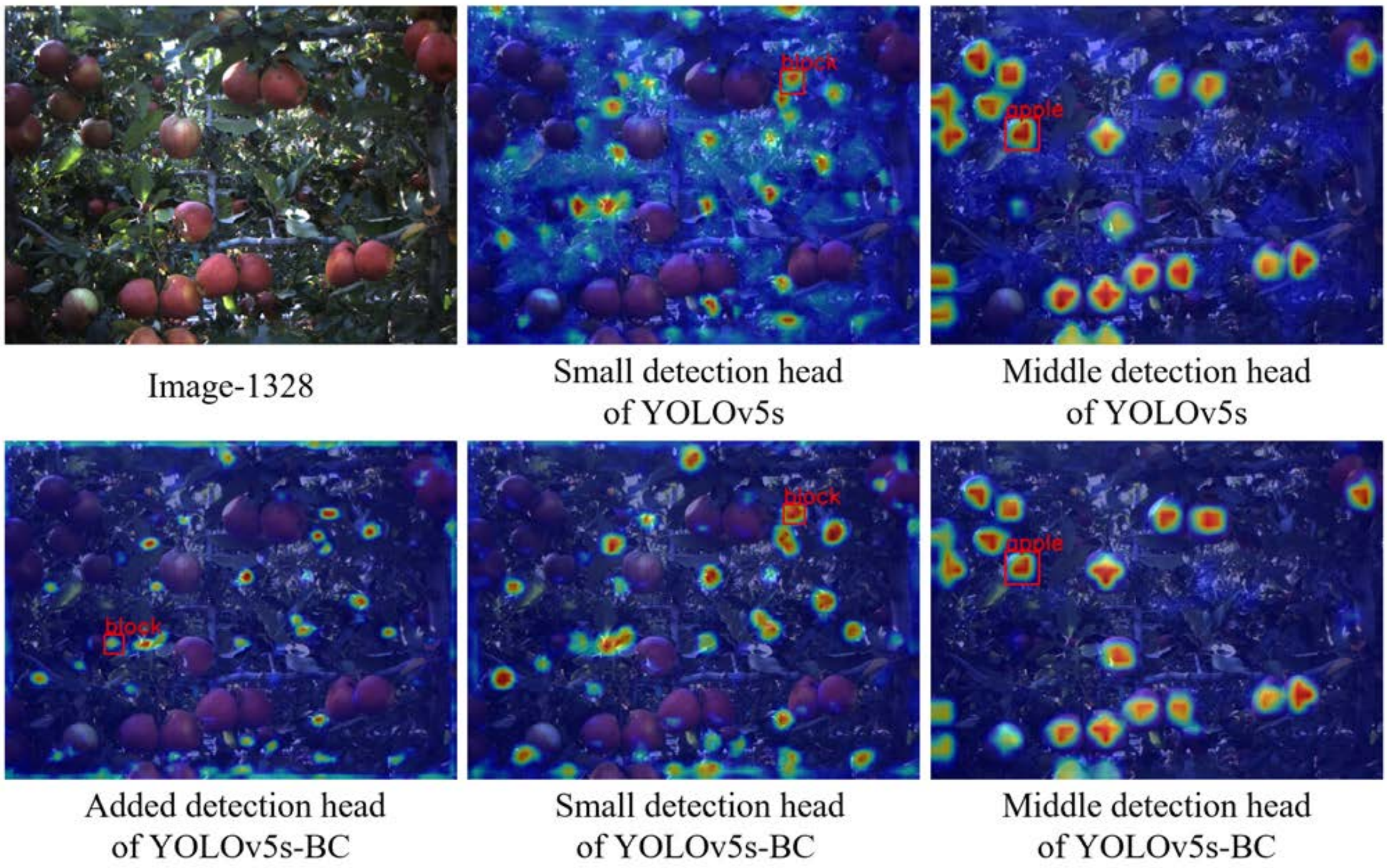

**Fig 11.** Heat maps of YOLOv5s-BC and YOLOv5s

To further test the generalization of the proposed model, new images were captured from a different apple orchard, as presented in Fig. 12. With the short-distance tests (see Fig. 12(a)), the model detects almost all apples accurately, and even those obstructed by leaves or tree branches are correctly recognized as block categories. Besides, in the long-distance tests, the detection of the model is also performed well. As shown in Fig. 12(b), it can detect all pickable and non-pickable apples on the whole tree. These two tests further demonstrate that our model can detect target apples in different apple orchards pretty well and has excellent generalization and robustness.

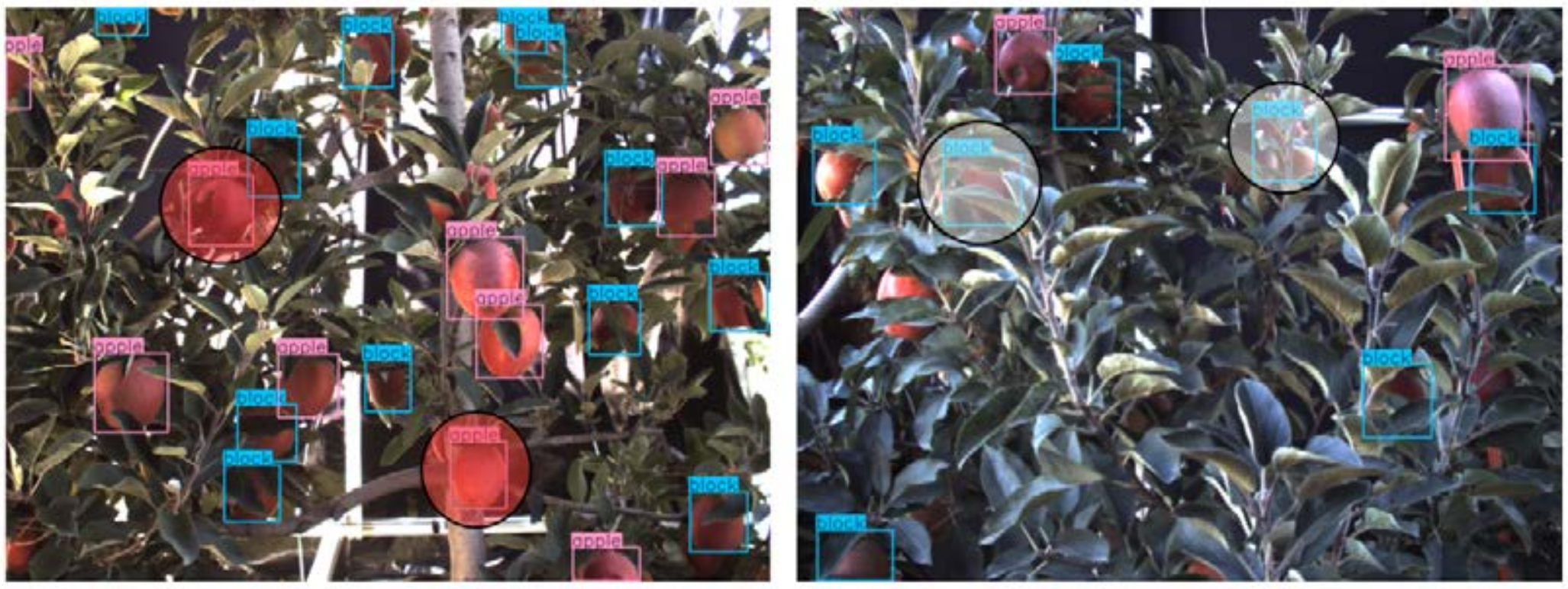

(a) Short distance

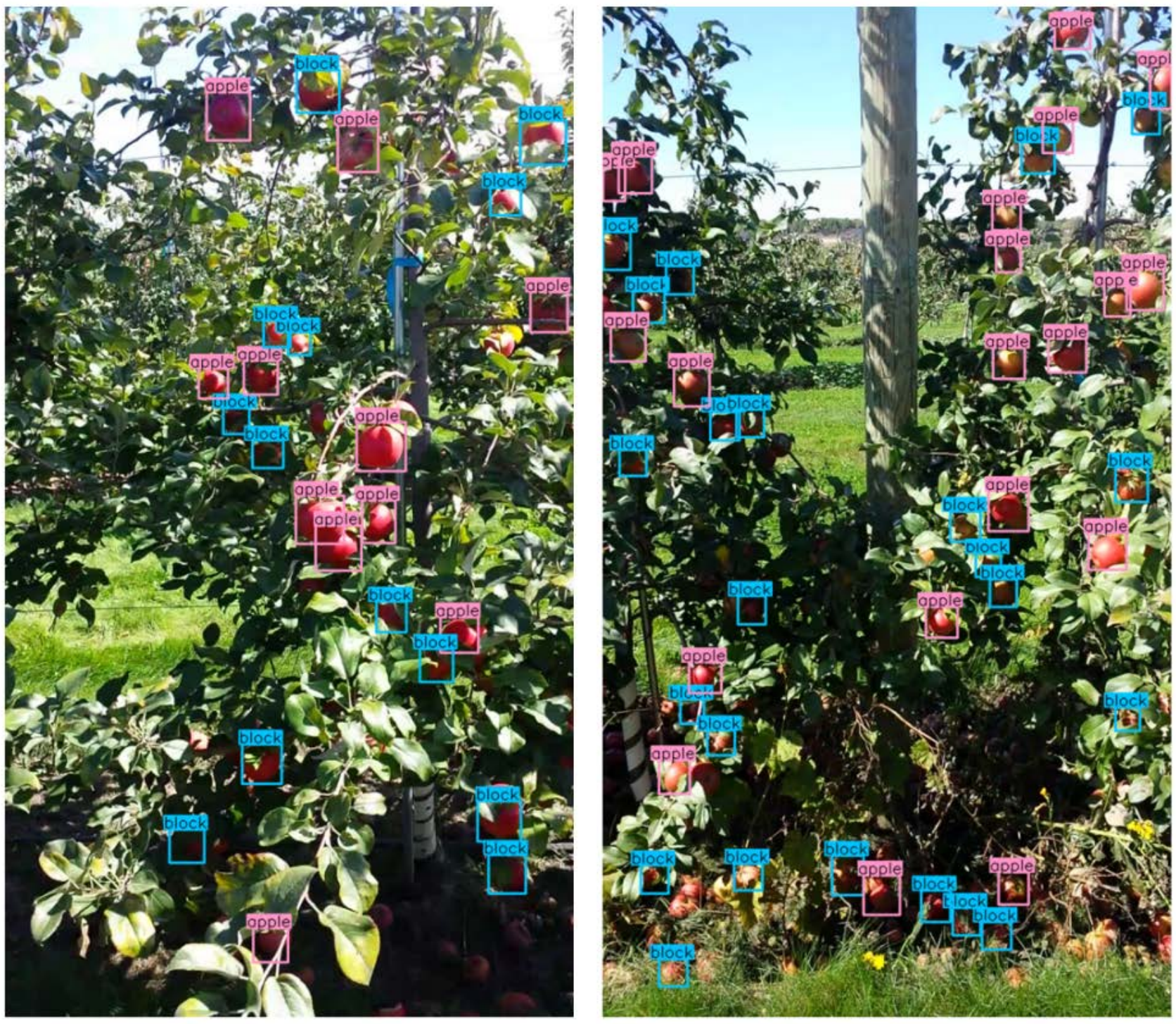

(b) Long distance

**Fig 12.** Model testing in different environments

## 5. Conclusions

In this paper, a real-time detection method based on YOLOv5s-BC is presented for apple detection. By adding a new detection head and combining the CA and BiFPN modules to optimize the YOLOv5s network model, the image features of target apples can be effectively extracted and the detection capability of smaller target apples can be enhanced. The detailed conclusions are summarized as follows.

The mAP performance of the YOLOv5-BC model on the test set reaches 88.7%, improving over the YOLOv5s, YOLOv4, YOLOv3, SSD, Faster R-CNN (ResNet50), and Faster R-CNN (VGG) models by 4.6%, 3.6%, 20.48%, 23.22%, 15.27%, and 15.59%. The weight size of the model is only 16.7 Mb, larger than the original YOLOv5s by 3 Mb, but smaller than YOLOv8, YOLOv4, YOLOv3, SSD, Faster R-CNN (ResNet), and Faster R-CNN (VGG) by 4.7, 239.3, 229.6, 74.4, 96.8, and 530.2 Mb. The detection of an image takes only 0.018 seconds, which guarantees the real-time requirements for apple detection. In the heat map, adding a new detection head to the model can detect apples from smaller targets. In addition, adding the CA mechanism makes the model pay more attention to and learn the high-level information of the detected targets and abandon other irrelevant information. In the test experiments at short and long distances, the proposed model can detect all the targets more perfectly, displaying the well robust performance of the model.

However, there are still certain limitations of the YOLOv5-BC model, such as the existence of a small number of missed or false detections. Therefore, the attention mechanism of the model needs to be further optimized, and the backbone network of the model needs to be modified as well, in order to further improve the detection accuracy of the proposed model.

**Data availability** The data that support the findings of this study are available from the corresponding author upon reasonable request.

## Declarations

**Conflict of interest** The authors declare that they have no conflict of interest.

## Acknowledgments

The present work is supported by the Start-up Funds from Wuhan University of Technology.

# References

Bao W. X., Zhu Z. Q., Hu G. S., et al. 2023. UAV remote sensing detection of tea leaf blight based on DDMA-YOLO. Computers and Electronics in Agriculture [J], 205 17. https://doi.org/10.1016/j.compag.2023.107637.

Bochkovskiy A., Wang C.-Y., Mark Liao H.-Y. 2020. YOLOv4: Optimal Speed and Accuracy of Object Detection. Arxiv [J]. https://doi.org/arXiv:2004.10934.

Fountas S., Mylonas N., Malounas I., et al. 2020. Agricultural Robotics for Field Operations. sensors [J], 20(9), 27. https://doi.org/10.3390/s20092672.

He K. M., Gkioxari G., Dollar P., et al. Mask R-CNN[C]//16th IEEE International Conference on Computer Vision (ICCV). Venice, ITALY:IEEE,2017:2980-2988.https://doi.org/10.1109/iccv.2017.322.

Hou Q., Zhou D., Feng J. Coordinate attention for efficient mobile network design[C]//Proceedings of the IEEE/CVF conference on computer vision and pattern recognition.2021:13713-13722.https://doi.org/10.48550/arXiv.2103.02907.

Jia W. K., Zhang Y., Lian J., et al. 2020. Apple harvesting robot under information technology: A review. International Journal of Advanced Robotic Systems [J], 17(3), 16. https://doi.org/10.1177/1729881420925310.

Li K. S., Wang J. C., Jalil H., et al. 2023. A fast and lightweight detection algorithm for passion fruit pests based on improved YOLOv5. Computers and Electronics in Agriculture [J], 204 11. https://doi.org/10.1016/j.compag.2022.107534.

Liang J. T., Chen X., Liang C. J., et al. 2023. A detection approach for late-autumn shoots of litchi based on unmanned aerial vehicle (UAV) remote sensing. Computers and Electronics in Agriculture [J], 204 10. https://doi.org/10.1016/j.compag.2022.107535.

Liu W., Anguelov D., Erhan D., et al. Ssd: Single shot multibox detector[C]//Computer Vision–

ECCV 2016: 14th European Conference, Amsterdam, The Netherlands, October 11–14, 2016, Proceedings, Part I 14.Springer,2016:21-37.https://doi.org/10.1007/978-3-319-46448-0_2.

Lu Y. Z., Young S. 2020. A survey of public datasets for computer vision tasks in precision agriculture. Computers and Electronics in Agriculture [J], 178 13. https://doi.org/10.1016/j.compag.2020.105760.

Lv J. D., Xu H., Han Y., et al. 2022. A visual identification method for the apple growth forms in the orchard. Computers and Electronics in Agriculture [J], 197 9. https://doi.org/10.1016/j.compag.2022.106954.

Qi J. T., Liu X. N., Liu K., et al. 2022. An improved YOLOv5 model based on visual attention mechanism: Application to recognition of tomato virus disease. Computers and Electronics in Agriculture [J], 194 12. https://doi.org/10.1016/j.compag.2022.106780.

Redmon J., Divvala S., Girshick R., et al. You Only Look Once: Unified, Real-Time Object Detection[C]//2016 IEEE Conference on Computer Vision and Pattern Recognition (CVPR). Seattle, WA:IEEE,2016:779-788.https://doi.org/10.1109/cvpr.2016.91.

Redmon J., Farhadi A. 2017. YOLO9000: Better, Faster, Stronger. IEEE Conference on Computer Vision & Pattern Recognition [J], 6517-6525. https://doi.org/10.1109/CVPR.2017.690.

Redmon J., Farhadi A. 2018. YOLOv3: An Incremental Improvement. Arxiv [J]. https://doi.org/arXiv:1804.02767.

Ren S. Q., He K. M., Girshick R., et al. 2017. Faster R-CNN: Towards Real-Time Object Detection with Region Proposal Networks. Ieee Transactions on Pattern Analysis and Machine Intelligence [J], 39(6), 1137-1149. https://doi.org/10.1109/tpami.2016.2577031.

Sun L. J., Hu G. R., Chen C., et al. 2022. Lightweight Apple Detection in Complex Orchards Using YOLOV5-PRE. Horticulturae [J], 8(12), 15. https://doi.org/10.3390/horticulturae8121169.

Tan M., Pang R., V. Le Q. 2020. EfficientDet: Scalable and Efficient Object Detection. Arxiv [J], 10778-10787. https://doi.org/arXiv:1911.09070.

Ultralytics yolov5 [M].

Wu F. Y., Duan J. L., Ai P. Y., et al. 2022. Rachis detection and three-dimensional localization of cut off point for vision-based banana robot. Computers and Electronics in Agriculture [J], 198 12. https://doi.org/10.1016/j.compag.2022.107079.

Xu B., Cui X., Ji W., et al. 2023. Apple Grading Method Design and Implementation for Automatic Grader Based on Improved YOLOv5. Agriculture-Basel [J], 13(1), 18. https://doi.org/10.3390/agriculture13010124.

Xu Z. B., Huang X. P., Huang Y., et al. 2022. A Real-Time Zanthoxylum Target Detection Method for an Intelligent Picking Robot under a Complex Background, Based on an Improved YOLOv5s Architecture. sensors [J], 22(2), 15. https://doi.org/10.3390/s22020682.

Yan B., Fan P., Lei X. Y., et al. 2021. A Real-Time Apple Targets Detection Method for Picking Robot Based on Improved YOLOv5. Remote sensing [J], 13(9), 23. https://doi.org/10.3390/rs13091619.

Yao J., Qi J. M., Zhang J., et al. 2021. A Real-Time Detection Algorithm for Kiwifruit Defects Based on YOLOv5. Electronics [J], 10(14), 13. https://doi.org/10.3390/electronics10141711.

Zhang D. Y., Luo H. S., Wang D. Y., et al. 2022. Assessment of the levels of damage caused by Fusarium head blight in wheat using an improved YoloV5 method. Computers and Electronics in Agriculture [J], 198 16. https://doi.org/10.1016/j.compag.2022.107086.

Zhao Y. S., Gong L., Huang Y. X., et al. 2016. A review of key techniques of vision-based control for harvesting robot. Computers and Electronics in Agriculture [J], 127 311-323. https://doi.org/10.1016/j.compag.2016.06.022.

Zheng Z. H., Wang P., Liu W., et al. Distance-IoU Loss: Faster and Better Learning for Bounding Box Regression[C]//34th AAAI Conference on Artificial Intelligence / 32nd Innovative Applications of Artificial Intelligence Conference / 10th AAAI Symposium on Educational Advances in Artificial Intelligence. New York, NY:Assoc Advancement Artificial Intelligence,2020:12993-13000.https://doi.org/10.48550/arXiv.1911.08287.

Zhou H. Y., Wang X., Au W., et al. 2022. Intelligent robots for fruit harvesting: recent developments and future challenges. Precision Agriculture [J], 23(5), 1856-1907. https://doi.org/10.1007/s11119-022-09913-3.

Zhu X. K., Lyu S. C., Wang X., et al. TPH-YOLOv5: Improved YOLOv5 Based on Transformer Prediction Head for Object Detection on Drone-captured Scenarios[C]//18th IEEE/CVF International Conference on Computer Vision (ICCV). Electr Network:Ieee Computer Soc,2021:2778-2788.https://doi.org/10.1109/iccvw54120.2021.00312.